\documentclass[11pt,twoside]{article}
\usepackage{asp2010}

\resetcounters

\begin{document}

\title{Dust production from sub-solar to super-solar metallicity in Thermally Pulsing
Asymptotic Giant Branch stars}
\author{Nanni Ambra$^{1,2}$, Bressan Alessandro$^1$, Marigo Paola$^2$, Girardi L\'eo$^3$, Javadi Atefeh$^4$ and van Loon Jacco$^5$
\affil{$^1$SISSA, Via Bonomea 265, I-34136 Trieste, Italy}
\affil{$^2$Dipartimento di Fisica e Astronomia Galileo Galilei,  Universit\`a di Padova, Vicolo dell'Osservatorio 3, I-35122 Padova, Italy}
\affil{$^3$Osservatorio Astronomico di Padova, Vicolo dell'Osservatorio 5, I-35122 Padova, Italy}
\affil{$^4$School of Astronomy, Institute for Research in Fundamental Sciences (IPM), 19395-5531 Tehran, Iran}
\affil{$^5$	Keele University, Keele, Staffordshire ST55BG, United Kingdom}}

\begin{abstract}
We discuss the dust chemistry and growth in the circumstellar envelopes (CSEs) of Thermally Pulsing Asymptotic Giant Branch (TP-AGB) star models computed with the \texttt{COLIBRI} code,
at varying initial mass and metallicity (Z=0.001, 0.008, 0.02, 0.04, 0.06).

A relevant result of our analysis deals with
the silicate production in M-stars. We show that, in order to reproduce the observed trend
between terminal velocities and mass-loss rates in Galactic M-giants, one has to significantly reduce the efficiency of chemisputtering by H$_2$
molecules, usually considered as
the most effective dust destruction mechanism. This indication is also in agreement with
the most recent laboratory results, which show that silicates may condense already at
T$_{\rm cond}\sim$1400~K, instead than at T$_{\rm cond}\sim$1000~K, as obtained by models that include
chemisputtering.

From the analysis of the total dust ejecta, we find that the total dust-
to-gas ejecta of intermediate-mass stars are much less dependent on metallicity than
usually assumed. In a broader context, our results are suitable to study the dust enrichment of the interstellar medium provided by TP-AGB stars in both nearby and high redshift
galaxies.
\end{abstract}

\section{Introduction}
During the TP-AGB phase, stars lose their envelopes at high rates,
typically between $10^{-8}$~M$_{\odot}$~yr$^{-1}$ and few $10^{-5}$~M$_{\odot}$~yr$^{-1}$, polluting the
surrounding Interstellar Medium (ISM) with metals partially condensed
into dust particles.

The metal abundances in the stellar atmospheres of TP-AGB stars,
 are crucially dependent on the stellar evolution at a given mass and metallicity.
In particular, the interplay between several physical processes occurring during this
complex phase, i.e. third dredge-up and hot bottom burning (HBB), determines the value of the C/O ratio, responsible
for the abrupt variation in the dust chemistry in the CSE. In fact, in case of C/O~$<1$, M-stars, the dust chemistry is oxygen-based, and the most abundant dust species are silicates (which possible inclusion of impurities), Al$_2$O$_3$ and iron, on the other hand if C/O~$>1$, C-stars, the chemistry is carbon-based, and the predominant dust species are amorphous carbon, silicon carbide and iron.

Once the bulk of dust is formed, the radiation pressure onto the dust grains can, under specific conditions, accelerate the particles, triggering a dust-driven wind.
If for C-stars it is now commonly accepted that amorphous carbon is the wind-driving species, for M-stars the details of the wind acceleration are still unclear and the problem has challenged several theoretical investigations \citep{Ireland06, Woitke06, Hofner09}.
In fact, in order to solve this problem, it is required that either large iron-free grains are produced, or that dirty silicate grains condense at higher temperatures than the one usually assumed in the literature \citep{Hofner_size08, Nanni13}.

In a broader context, the dust produced by TP-AGB stars enters the life cycle of dust in galaxies, representing one of the main stellar dust sources both in the Milky Way and in high redshifts galaxies, where the large dust reservoirs observed could be mainly ascribed to the ejecta of TP-AGB stars more massive than 3~M$_\odot$ \citep{Gehrz89, Dwek07, Valiante09, Dwek11}.

\section{The model}
Following \citet{GS99, FG06}, we studied the problem of dust formation coupled with a stationary, spherically symmetric wind.
The model requires, as input parameters, some important stellar quantities that vary along the TP-AGB tracks, i.e. stellar mass, luminosity, effective temperature, elemental abundances in the atmosphere (C/O ratio) and mass-loss rate.

The TP-AGB tracks used in these computations, are calculated with the \texttt{COLIBRI} code by \citet{marigoetal13} with the revised mass-loss rates described in \citet{rosenfield14}. The initial conditions taken by \texttt{COLIBRI} are provided by the output of the most recent evolutionary tracks computed with the Padova-Trieste code \texttt{PARSEC} \citep{Bressanetal12}.
We remind that the \texttt{COLIBRI} code includes an on-the-fly computation of the detailed chemistry of 800 molecular species and opacities \citep{MarigoAringer_09}. The HBB is described by a complete network of nuclear reactions (CNO, NeNa, MgAl cycles) coupled with a diffuse description of convection. The mass-loss rate and third dredge-up are given by parameterized description.

In our dust formation scheme, fully described in \citet{Nanni13, Nanni14}, the dust growth is computed from the balance between two terms: the accretion and the destruction rates. Different destruction mechanisms can be at work. In the standard approach, the reaction between H$_2$ molecules and the silicate grain surface, known as ``chemisputtering'', turns out to be much more efficient than pure sublimation induced by the stellar radiation heating, so that only the former process is usually included in the models \citep{GS99, FG06, Ventura12}.
\section{Results and discussion}
\subsection{Silicate condensation temperature in CSEs of M-stars}
\begin{figure*}
\centering
\includegraphics[width=0.45\textwidth]{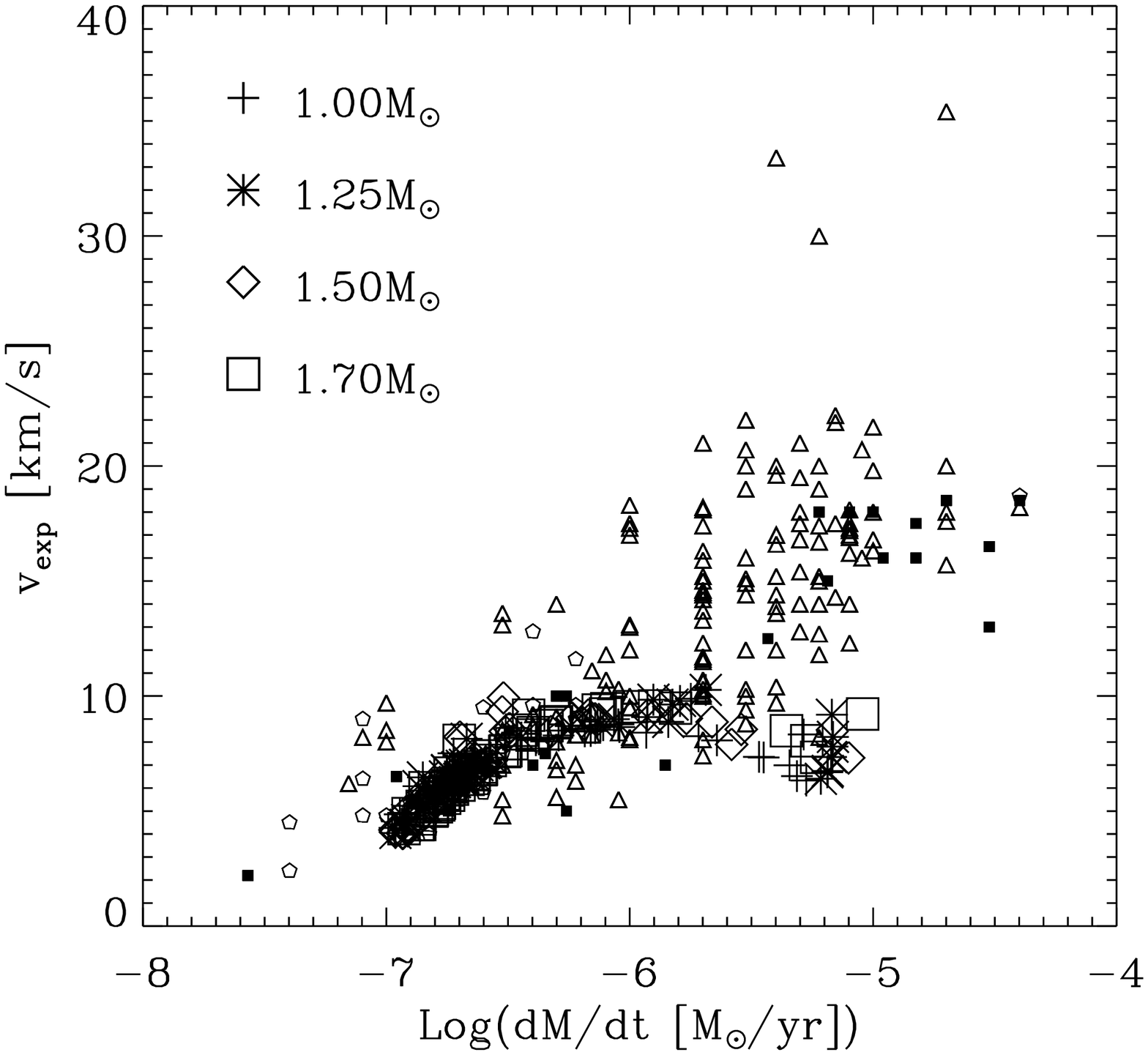}
\includegraphics[width=0.45\textwidth]{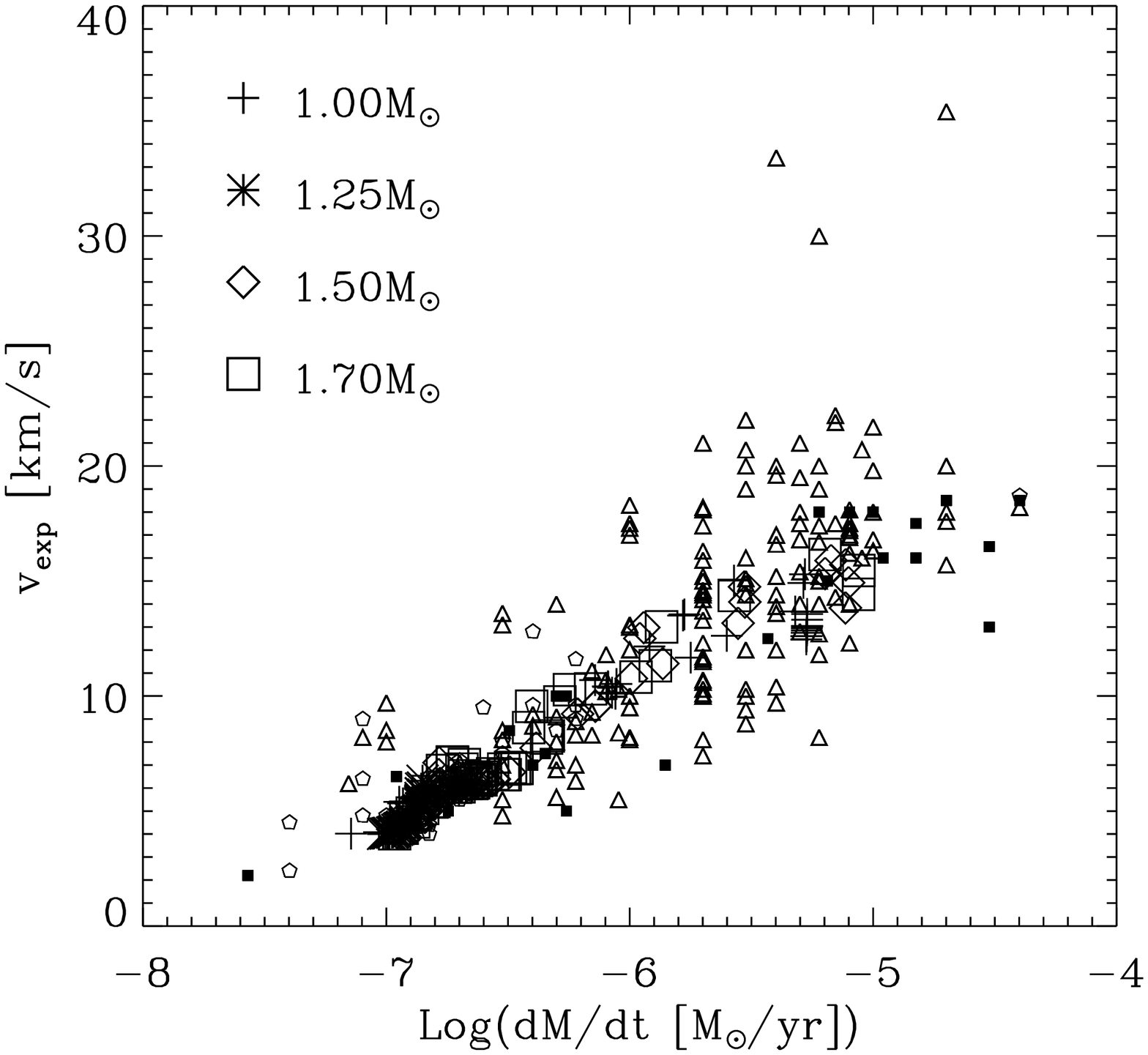}
\caption{Expansion velocities against mass-loss rates in CSEs
of variable M-stars. Observations of Galactic M-stars by \citet{Loup93} (triangles),
\citet{Gonzalez03} (pentagons) and \citet{Schoier13} (filled squares)
are compared with predicted expansion velocities for a few selected TP-AGB tracks with $Z=0.02$
for the values of initial stellar masses listed in upper left of each figure.
\textit{Left panel}: models in which the chemisputtering process is fully efficient.
\textit{Right panel}: models in which the chemisputtering process is fully inhibited (only sublimation).}
\label{v_dotm}
\end{figure*}
We first considered in our models fully efficient chemisputtering and dirty silicates opacities by \citet{Ossenkopf92}. We compared the predicted expansion velocities as a function of the mass-loss rate, for different stellar masses at Z=0.02, with Galactic data. The results are shown in the left panel of Fig.~\ref{v_dotm}.
From this figure it is clear that, by including chemisputtering in our model, the observed trend is not reproduced. In particular, the velocity at a given value of mass-loss rate is underestimated at high mass-loss rates.

This discrepancy is not exclusively present in our models, but is a long standing problem that has challenged many theoretical investigations \citep{Ireland06, Woitke06, Hofner09}.
Such a discrepancy is resolved if silicates can condense close enough to the stellar photosphere ($r\sim2$~$R_*$), where the acceleration of gravity is sufficiently large to produce the final expansion velocities.

One possible way of obtaining this condition consists of neglecting the chemisputtering process in our calculations.
Contrary to pure sublimation, in fact, the chemisputtering efficiency remains high even at low temperatures producing a significant delay in the condensation of silicates ($r\sim5$~$R_*$).

We therefore revised the efficiency of the chemisputtering process by taking into account the experimental studies on the kinetics of silica erosion in presence of H$_2$ gas. The results of these experiments yield an activation energy barrier of the chemisputtering reaction too high to be efficient at the typical ranges of temperature and pressure of the CSEs \citep[Tielens, private communication]{Gardner74,Tso82}.
 These results are in agreement with the ones performed by \citet{Nagahara96} in which the evaporation of silicates in presence of H$_2$ molecules is only relevant at pressures at least two orders of magnitude higher
 than the ones found in these CSEs at temperatures of 2000~K.

 From these considerations, we therefore computed a set of models in which only pure sublimation is included and chemisputtering is totally neglected.
 As expected, in this case the condensation temperature of silicates is significantly higher, T$_{\rm cond}\sim$1400~K, than the one obtained when chemisputtering is taken into account, T$_{\rm cond}\sim$1000~K \citep{Nanni13}. Remarkably, also in recent experiments in which the condensation process of silicate dust is studied, the condensation temperatures measured can be up to T$_{\rm cond}\sim$1400~K \citep{Nagahara09}.

 The results of the predicted expansion velocities as a function of the mass-loss rate for these new computations are shown in the right panel of Fig.~\ref{v_dotm}, from which it is clear that the trend obtained is now fairly well reproduced.

\subsection{Total dust ejecta and dust-to-gas ratios}
\label{ejecta}

\begin{figure*}
\centering
\includegraphics[width=0.45\textwidth]{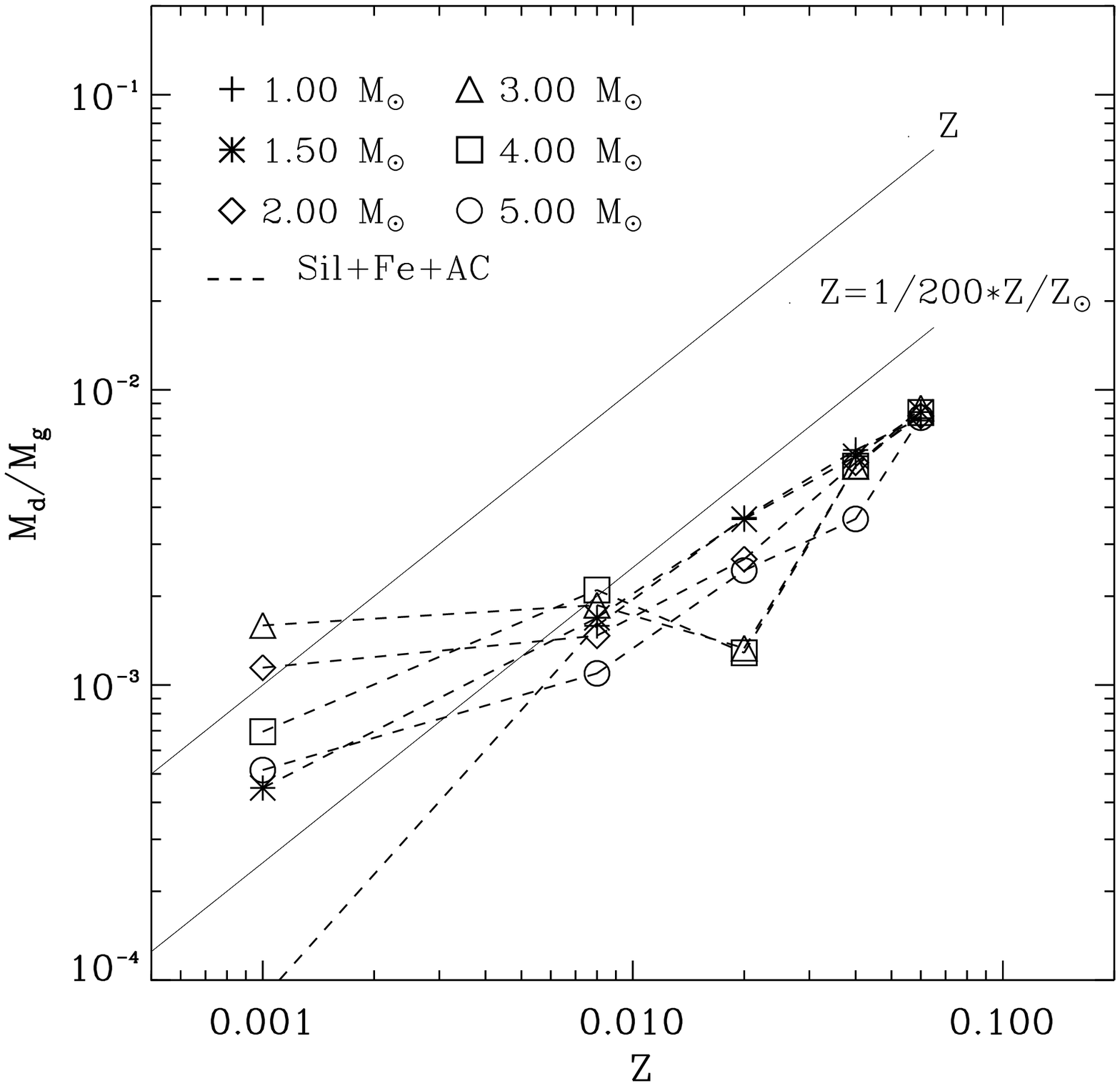}
\caption{Total dust-to-gas ejecta as a function of
the initial metallicity and for different stellar masses. The computations are based on the TP-AGB tracks by \citet{marigoetal13} with the new mass-loss prescription by \citet{rosenfield14}. The dust formation scheme is described in \citet{Nanni13, Nanni14}.}
\label{d2g}
\end{figure*}

We calculate the dust ejecta produced during the entire duration of the TP-AGB phase for stars of various masses and in a wide range of metallicity
(from Z=0.001 to Z=0.06).

The wide range of metallicity considered allows us to study the trend of the total dust-to-gas ratio (total dust produced over the total gas released) as a function of the metallicity for different initial stellar masses.
The results are compared with two standard assumption adopted in the literature in the framework of galaxy evolution, as shown in Fig.~\ref{d2g}.

A very interesting result obtained through our investigation is that the total dust-to-gas ratio of intermediate-mass stars is less dependent on the metallicity than what it is usually assumed. More striking, at the lowest metallicity, the amount of metals condensed into dust (mostly carbon) can be higher than the initial metallicity, due to the efficient third dredge-up that brings carbon up to the stellar surface \citep{Nanni14}.

\section{Summary and conclusions}
We revised the efficiency of the chemisputtering process, usually assumed to be the most important destruction mechanism acting on silicate grains.
In particular, the comparison between our predictions and observations provided an important indication about the fact that
this process can be highly inhibited in the typical conditions of these CSEs.
This indication turns out to be in agreement with experimental results about the evaporation and condensation of silicates.

The model developed can be employed to compute the total dust-to-gas ratio as a function of the initial metallicity, especially relevant for the study of dust evolution in galaxies. In particular, we found a milder dependence of the dust-to-gas ratio on the initial metallicity than the one usually assumed in the literature.
\acknowledgements
We acknowledge the support from Progetto di Ateneo 2012 (University of Padova, ID: CPDA125588/12), and from
ERC Consolidator Grant (ID: 610654, project STARKEY).
\bibliography{nanni}

\begin{thebibliography}{}
\expandafter\ifx\csname natexlab\endcsname\relax\def\natexlab#1{#1}\fi
\expandafter\ifx\csname url\endcsname\relax
  \def\url#1{\texttt{#1}}\fi
\expandafter\ifx\csname urlprefix\endcsname\relax\def\urlprefix{URL }\fi
\providecommand{\eprint}[2][]{\url{#2}}

\bibitem[{{Bressan} et~al.(2012){Bressan}, {Marigo}, {Girardi}, {Salasnich},
  {Dal Cero}, {Rubele}, \& {Nanni}}]{Bressanetal12}
{Bressan}, A., {Marigo}, P., {Girardi}, L., {Salasnich}, B., {Dal Cero}, C.,
  {Rubele}, S., \& {Nanni}, A. 2012, \mnras, 427, 127. \eprint{1208.4498}

\bibitem[{{Dwek} \& {Cherchneff}(2011)}]{Dwek11}
{Dwek}, E., \& {Cherchneff}, I. 2011, \apj, 727, 63. \eprint{1011.1303}

\bibitem[{{Dwek} et~al.(2007){Dwek}, {Galliano}, \& {Jones}}]{Dwek07}
{Dwek}, E., {Galliano}, F., \& {Jones}, A.~P. 2007, Nuovo Cimento B Serie, 122,
  959

\bibitem[{{Ferrarotti} \& {Gail}(2006)}]{FG06}
{Ferrarotti}, A.~S., \& {Gail}, H.-P. 2006, \aap, 447, 553

\bibitem[{{Gail} \& {Sedlmayr}(1999)}]{GS99}
{Gail}, H.-P., \& {Sedlmayr}, E. 1999, \aap, 347, 594

\bibitem[{{Gardner}(1974)}]{Gardner74}
{Gardner}, R.~A. 1974, Journal of solid state chemistry, 9, 336

\bibitem[{{Gehrz}(1989)}]{Gehrz89}
{Gehrz}, R. 1989, in Interstellar Dust, edited by L.~J. {Allamandola}, \&
  A.~G.~G.~M. {Tielens}, vol. 135 of IAU Symposium, 445

\bibitem[{{Gonz{\'a}lez Delgado} et~al.(2003){Gonz{\'a}lez Delgado},
  {Olofsson}, {Kerschbaum}, {Sch{\"o}ier}, {Lindqvist}, \&
  {Groenewegen}}]{Gonzalez03}
{Gonz{\'a}lez Delgado}, D., {Olofsson}, H., {Kerschbaum}, F., {Sch{\"o}ier},
  F.~L., {Lindqvist}, M., \& {Groenewegen}, M.~A.~T. 2003, \aap, 411, 123.
  \eprint{astro-ph/0302179}

\bibitem[{{H{\"o}fner}(2008)}]{Hofner_size08}
{H{\"o}fner}, S. 2008, Physica Scripta Volume T, 133, 014007

\bibitem[{{H{\"o}fner}(2009)}]{Hofner09}
--- 2009, in Cosmic Dust - Near and Far, edited by T.~{Henning}, E.~{Gr{\"u}n},
  \& J.~{Steinacker}, vol. 414 of Astronomical Society of the Pacific
  Conference Series, 3. \eprint{0903.5280}

\bibitem[{{Ireland} \& {Scholz}(2006)}]{Ireland06}
{Ireland}, M.~J., \& {Scholz}, M. 2006, \mnras, 367, 1585.
  \eprint{arXiv:astro-ph/0601383}

\bibitem[{{Loup} et~al.(1993){Loup}, {Forveille}, {Omont}, \& {Paul}}]{Loup93}
{Loup}, C., {Forveille}, T., {Omont}, A., \& {Paul}, J.~F. 1993, \aaps, 99, 291

\bibitem[{{Marigo} \& {Aringer}(2009)}]{MarigoAringer_09}
{Marigo}, P., \& {Aringer}, B. 2009, A\&A, 508, 1539. \eprint{0907.3248}

\bibitem[{{Marigo} et~al.(2013){Marigo}, {Bressan}, {Nanni}, {Girardi}, \&
  {Pumo}}]{marigoetal13}
{Marigo}, P., {Bressan}, A., {Nanni}, A., {Girardi}, L., \& {Pumo}, M.~L. 2013,
  \mnras, 434, 488. \eprint{1305.4485}

\bibitem[{{Nagahara} et~al.(2009){Nagahara}, {Ogawa}, {Ozawa}, {Tamada},
  {Tachibana}, \& {Chiba}}]{Nagahara09}
{Nagahara}, H., {Ogawa}, R., {Ozawa}, K., {Tamada}, S., {Tachibana}, S., \&
  {Chiba}, H. 2009, in Cosmic Dust - Near and Far, edited by T.~{Henning},
  E.~{Gr{\"u}n}, \& J.~{Steinacker}, vol. 414 of Astronomical Society of the
  Pacific Conference Series, 403

\bibitem[{{Nagahara} \& {Ozawa}(1996)}]{Nagahara96}
{Nagahara}, H., \& {Ozawa}, K. 1996, Geochim.~Cosmochim.~Acta, 60, 1445

\bibitem[{{Nanni} et~al.(2013){Nanni}, {Bressan}, {Marigo}, \&
  {Girardi}}]{Nanni13}
{Nanni}, A., {Bressan}, A., {Marigo}, P., \& {Girardi}, L. 2013, \mnras, 434,
  2390. \eprint{1306.6183}

\bibitem[{{Nanni} et~al.(2014){Nanni}, {Bressan}, {Marigo}, \&
  {Girardi}}]{Nanni14}
--- 2014, \mnras. \eprint{1312.0875}

\bibitem[{{Ossenkopf} et~al.(1992){Ossenkopf}, {Henning}, \&
  {Mathis}}]{Ossenkopf92}
{Ossenkopf}, V., {Henning}, T., \& {Mathis}, J.~S. 1992, \aap, 261, 567

\bibitem[{{Rosenfield} et~al.(2014){Rosenfield}, {Marigo}, {Girardi},
  {Dalcanton}, {Bressan}, {Gullieuszik}, {Weisz}, {Williams}, {Dolphin}, \&
  {Aringer}}]{rosenfield14}
{Rosenfield}, P., {Marigo}, P., {Girardi}, L., {Dalcanton}, J.~J., {Bressan},
  A., {Gullieuszik}, M., {Weisz}, D., {Williams}, B.~F., {Dolphin}, A., \&
  {Aringer}, B. 2014, \apj, 790, 22. \eprint{1406.0676}

\bibitem[{{Sch{\"o}ier} et~al.(2013){Sch{\"o}ier}, {Ramstedt}, {Olofsson},
  {Lindqvist}, {Bieging}, \& {Marvel}}]{Schoier13}
{Sch{\"o}ier}, F.~L., {Ramstedt}, S., {Olofsson}, H., {Lindqvist}, M.,
  {Bieging}, J.~H., \& {Marvel}, K.~B. 2013, \aap, 550, A78. \eprint{1301.2129}

\bibitem[{{Tso} \& {Pask}(1982)}]{Tso82}
{Tso}, S.~T., \& {Pask}, J.~A. 1982, Journal of the American Ceramic Society,
  65, 457

\bibitem[{{Valiante} et~al.(2009){Valiante}, {Schneider}, {Bianchi}, \&
  {Andersen}}]{Valiante09}
{Valiante}, R., {Schneider}, R., {Bianchi}, S., \& {Andersen}, A.~C. 2009,
  \mnras, 397, 1661. \eprint{0905.1691}

\bibitem[{{Ventura} et~al.(2012){Ventura}, {Criscienzo}, {Schneider}, {Carini},
  {Valiante}, {D'Antona}, {Gallerani}, {Maiolino}, \&
  {Tornamb{\'e}}}]{Ventura12}
{Ventura}, P., {Criscienzo}, M.~D., {Schneider}, R., {Carini}, R., {Valiante},
  R., {D'Antona}, F., {Gallerani}, S., {Maiolino}, R., \& {Tornamb{\'e}}, A.
  2012, \mnras, 424, 2345. \eprint{1205.6216}

\bibitem[{{Woitke}(2006)}]{Woitke06}
{Woitke}, P. 2006, \aap, 460, L9. \eprint{arXiv:astro-ph/0609392}

\end{thebibliography}
\end{document}